\documentclass[epj]{webofc}
\usepackage[varg]{txfonts}   
\usepackage{lineno}
\begin{document}
%
%
\title{HAWC High Energy Upgrade with a Sparse Array}
%
%

\author{V.Joshi\inst{1}\fnsep\thanks{\email{vikas.joshi@mpi-hd.mpg.de}} 
        for the HAWC collaboration\inst{2}
}

\institute{Max Planck Institut f\"{u}r Kernphysik, Heidelberg, Germany
\and
           For a complete author list, see www.hawc-observatory.org/collaboration/
          }

\abstract{%
The High Altitude Water Cherenkov (HAWC) gamma-ray observatory has been fully operational since March 2015. To improve its sensitivity at the highest energies, it is being upgraded with an additional sparse array called outrigger array. We will discuss in this contribution, the different outrigger array components, and the simulation results to optimize it. }
\maketitle
\section{HAWC and the Motivation for Outriggers}
\label{intro}
HAWC is situated in central Mexico at an altitude of 4100 m above the sea level. It has a wide field of view of 2 sr and operational energy range of 0.1-100 TeV. It consists of 300 Water Cherenkov Detectors (WCDs) in the main array encompassing a surface area of 20000 m$^{2}$. The main array WCDs comprised of cylindrical steel water tanks of diameter 7.3 m and height 4.5 m with 4 Photo Multiplier Tubes (PMTs) (three 8” and one 10”) in each one of them. HAWC detects the Cherenkov light produced in the water by particles generated in an atmospheric air shower.

When the energy of the primary particle is of the order of tens of TeV, the footprint of the shower becomes comparable to the size of the main array. Therefore, most of the recorded showers are not contained in the array, which causes challenges to constrain the shower properties. To address these challenges the construction of the outrigger array around the main array has started. It will increase the fraction of well-reconstructed showers above multi-TeV energies. The outrigger array will help in determining the position of the core of the shower falling outside the main array and it will also improve the determination of the primary particle's direction and energy. 
\section{Outrigger Array}
\label{outrigger array}
\begin{figure}[ht]
\centering
\includegraphics[scale=0.16]{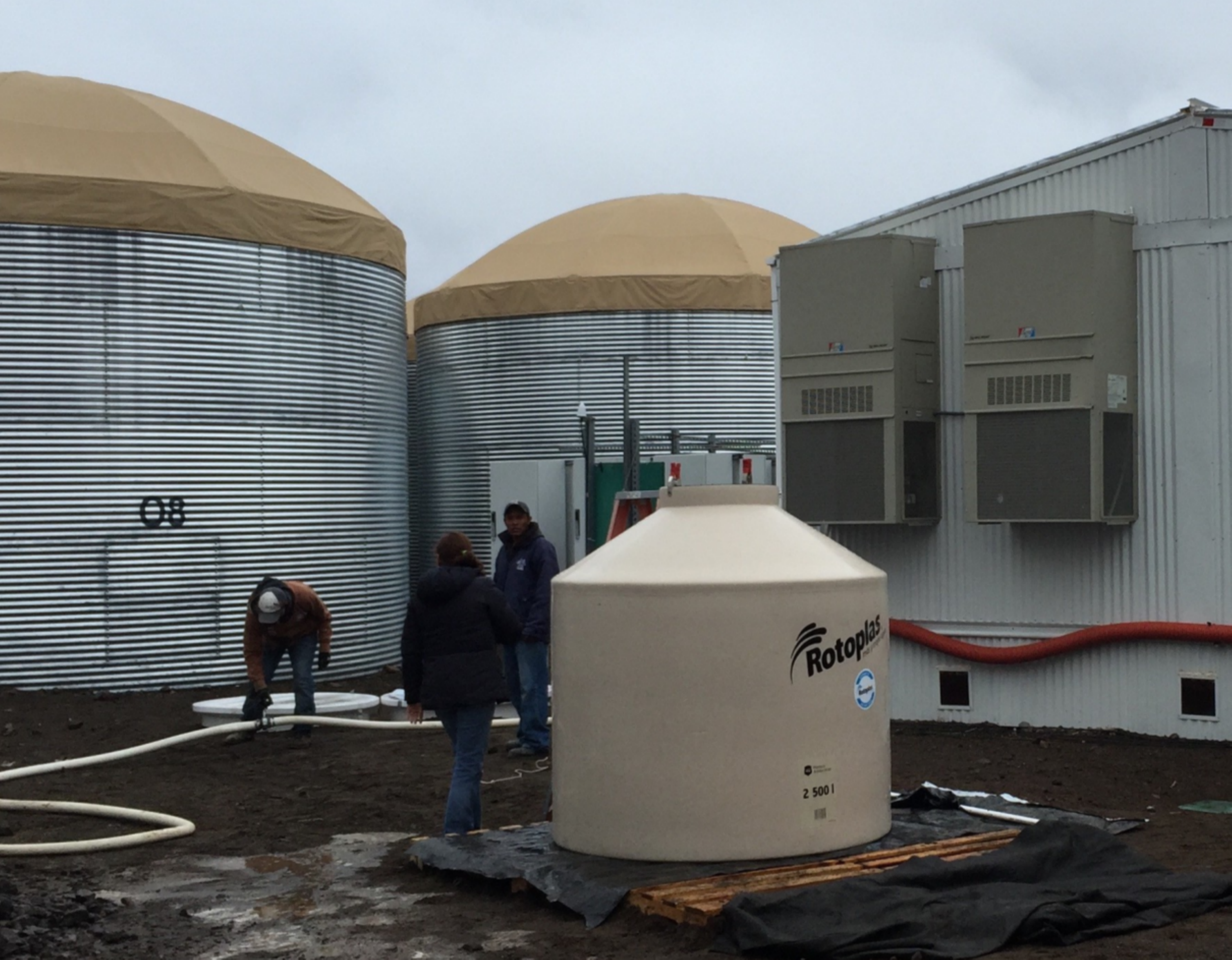}
\includegraphics[scale=0.18]{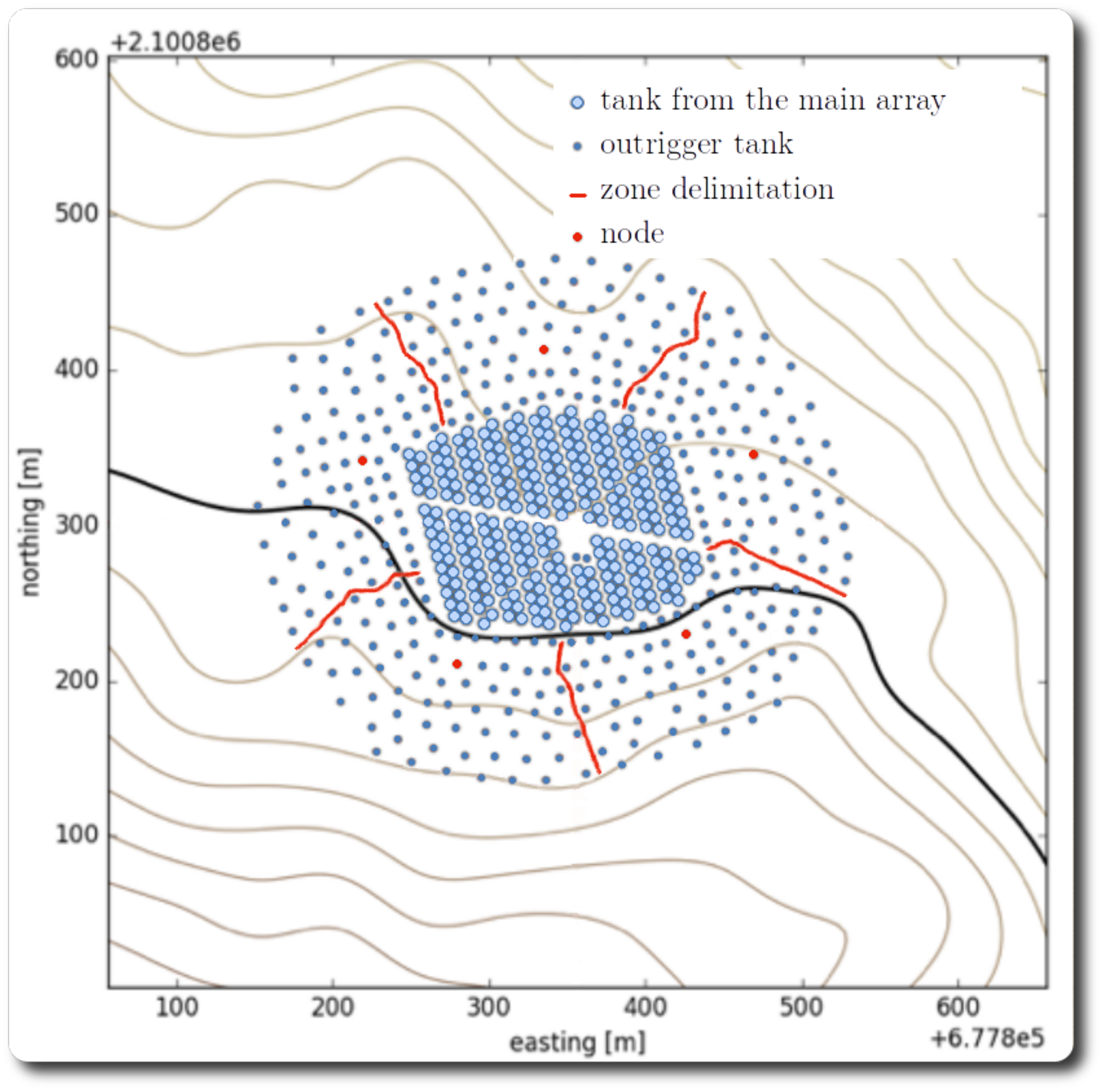}
\caption{\textbf{a.} Outrigger tank and main array tanks. \textbf{b.} Outrigger array surrounding the main HAWC array. The red lines shows the different sections of the outrigger array.}
\label{fig-1}       
\end{figure}

The outrigger array \cite{ref_outrigger} consists of 350 cylindrical tanks of diameter 1.55 m and height 1.65 m (see Figure~\ref{fig-1}a). Each tank has one Hamamatsu R5912 8" PMT at the bottom of the tank. The outrigger array will be deployed in a circular symmetric way around the main HAWC array with a mutual separation of 12 m to 18 m (see Figure~\ref{fig-1}b). 

To trigger and readout, the system electronics developed for the FlashCAM \cite{ref_flashcam} will be used. FlashCAM is a readout electronics, which has been developed for the cameras of the medium-size telescopes of the Cherenkov Telescope Array. The reason for using the FlashCam readout for outrigger array is that each PMT of the outrigger array is equivalent to a pixel of an Imaging Atmospheric Cherenkov Telescopes (IACTs) camera. The outrigger array is divided into five sections with 70 outriggers in each of them. One such section will contain a readout and trigger electronics, which we named as the: Flash Adc electronics for the Cherenkov Outrigger Node (FALCON). A node will contain 3 Flash-ADC boards, each of them can digitize 24 channels with a sampling speed of 250 MHz with a 12-bit accuracy. It also allows a flexible digital multiplicity trigger as well as the readout of full waveforms, with settable length (typically 40 samples i.e. 160 ns) which can be used for charge extraction and signal timing information.

\section{Simulations}
\label{simulations}
We performed extensive simulations in order to optimize the outrigger array. This can be further divided into two parts:

\begin{itemize}
\item[\textbf{1.}] Simulations to study the effect of different PMT options and tank colors.
\item[\textbf{2.}] Simulations to develop a likelihood fit method in order to fit the shower core and to constrain the shower energy and the depth of the shower maximum.
\end{itemize}

\subsection{Simulations for PMT Options and Tank Colors}
\label{simulatioins for PMT and tank}
In order to choose the size of the PMT, different PMT sizes have been simulated in combination with different tank wall colors. Here we present the results for the 3" and 8" PMT with tank wall colors black and white. We have focused on the following figures of merit:
\begin{itemize}
\item[\textbf{1.}] Average number of Photo-Electrons (PEs) observed at a given distance from the shower core.
\item[\textbf{2.}] RMS of the distribution of the time difference between neighboring tank pairs for the arrival time of the first PE.
\end{itemize}

\begin{figure}[ht]
\centering
\includegraphics[scale=0.35]{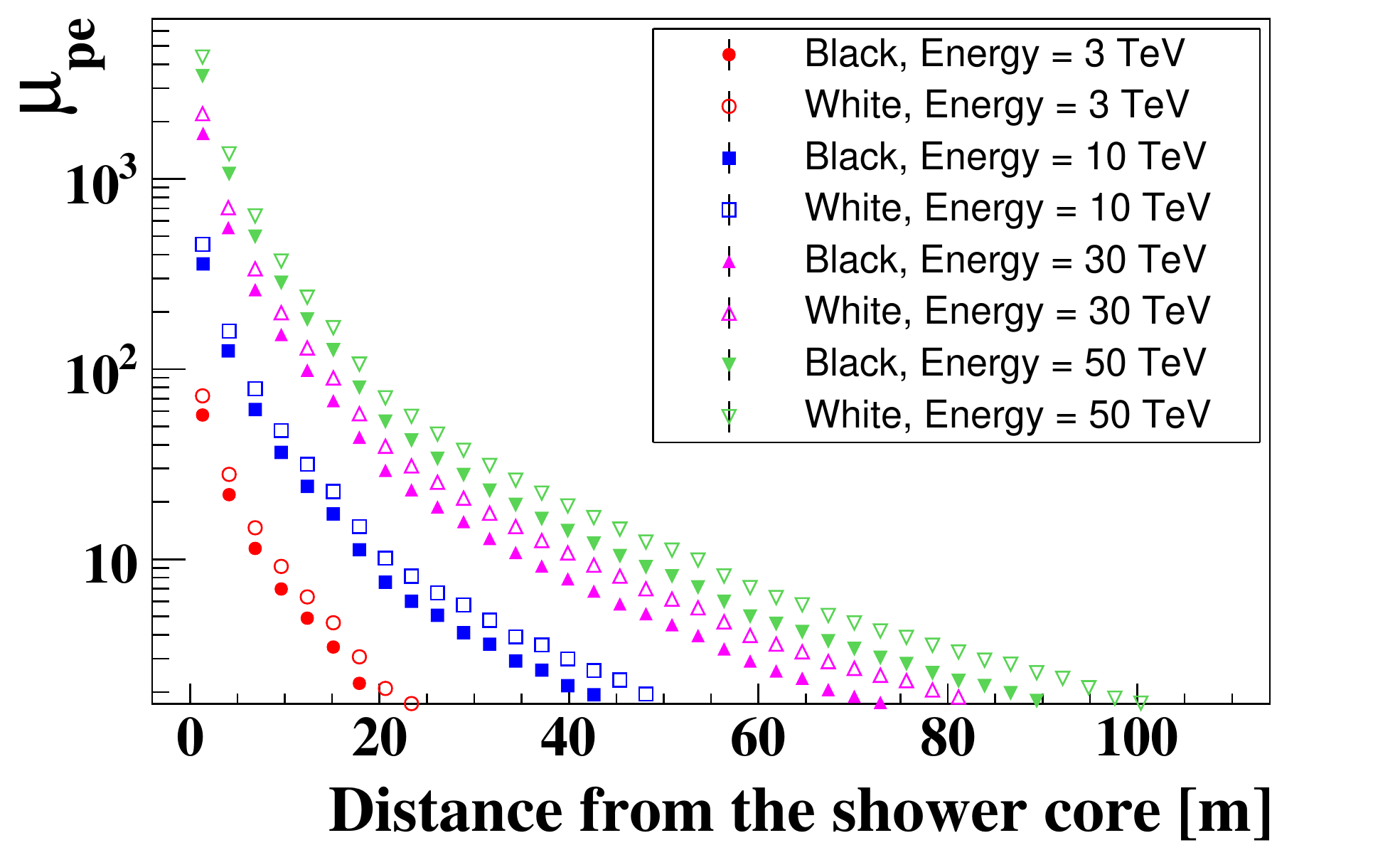}
\includegraphics[scale=0.35]{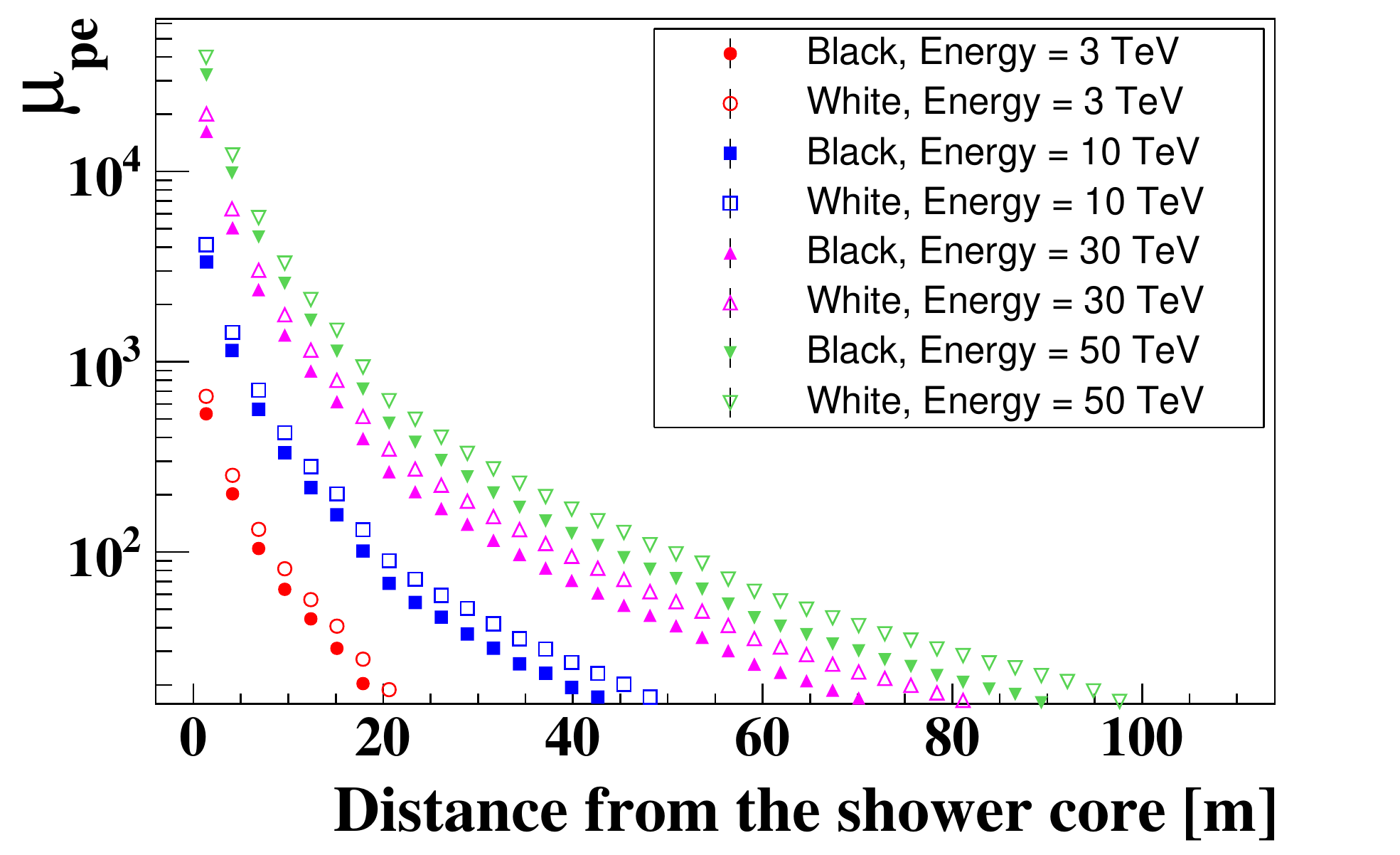}
\caption{Average number of PE ($\mu_{pe}$ ) observed for 3” PMT (left) and 8" PMT (right) with black and white tanks as a function of distance from the shower core for different energies.}
\label{fig-2}       
\end{figure}

\begin{figure}[ht]
\centering
\includegraphics[scale=0.45]{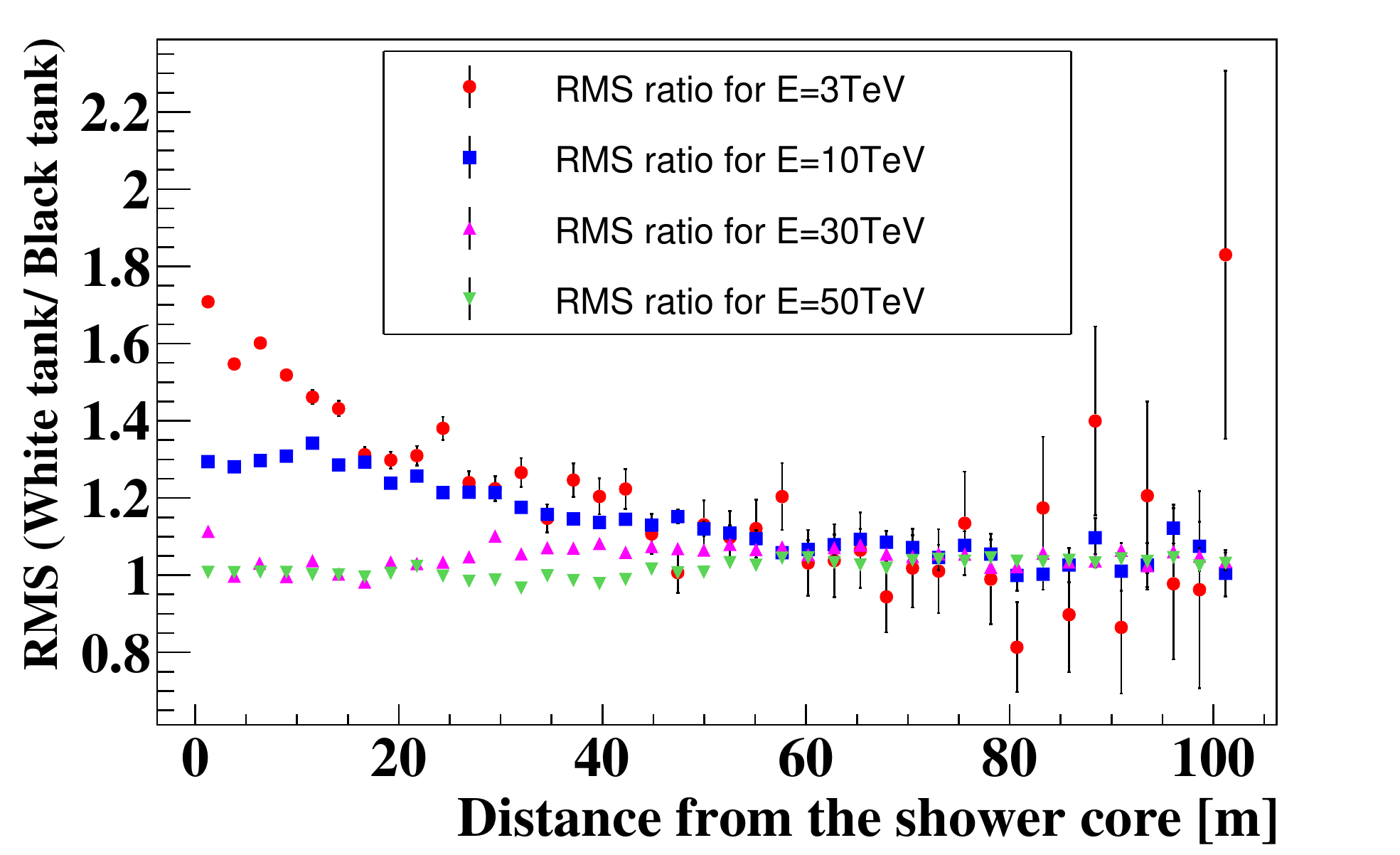}
\caption{Ratio of the RMSs (see Section~\ref{simulatioins for PMT and tank}) for white/black tanks for 8” PMT as a function of distance from the shower core for different energies (E).}
\label{fig-3}       
\end{figure}

It can be seen from the Figure~\ref{fig-2} that one gets 10 times more PEs with the 8" PMT in comparison to the 3" PMT and the effect of the white wall color in the contrast of black wall color is 20\% increase in the number of PEs observed. Furthermore, the white wall color is more diffusive than the black wall color, and the loss of the timing information by using the white wall color it can be more than 20\% (see Figure~\ref{fig-3}) in comparison to the black wall color. It can be concluded that we don't gain much in the average number of PEs observed by using the white wall color and we lose considerably in our timing information. We decided that black wall color tanks (less diffusive) with 8" PMT seems to be the appropriate choice.

\subsection{Simulations for Likelihood Core Fit Method}
\label{simulaitons for likelihood method}
\begin{figure}[ht]
\centering
\includegraphics[scale=0.54]{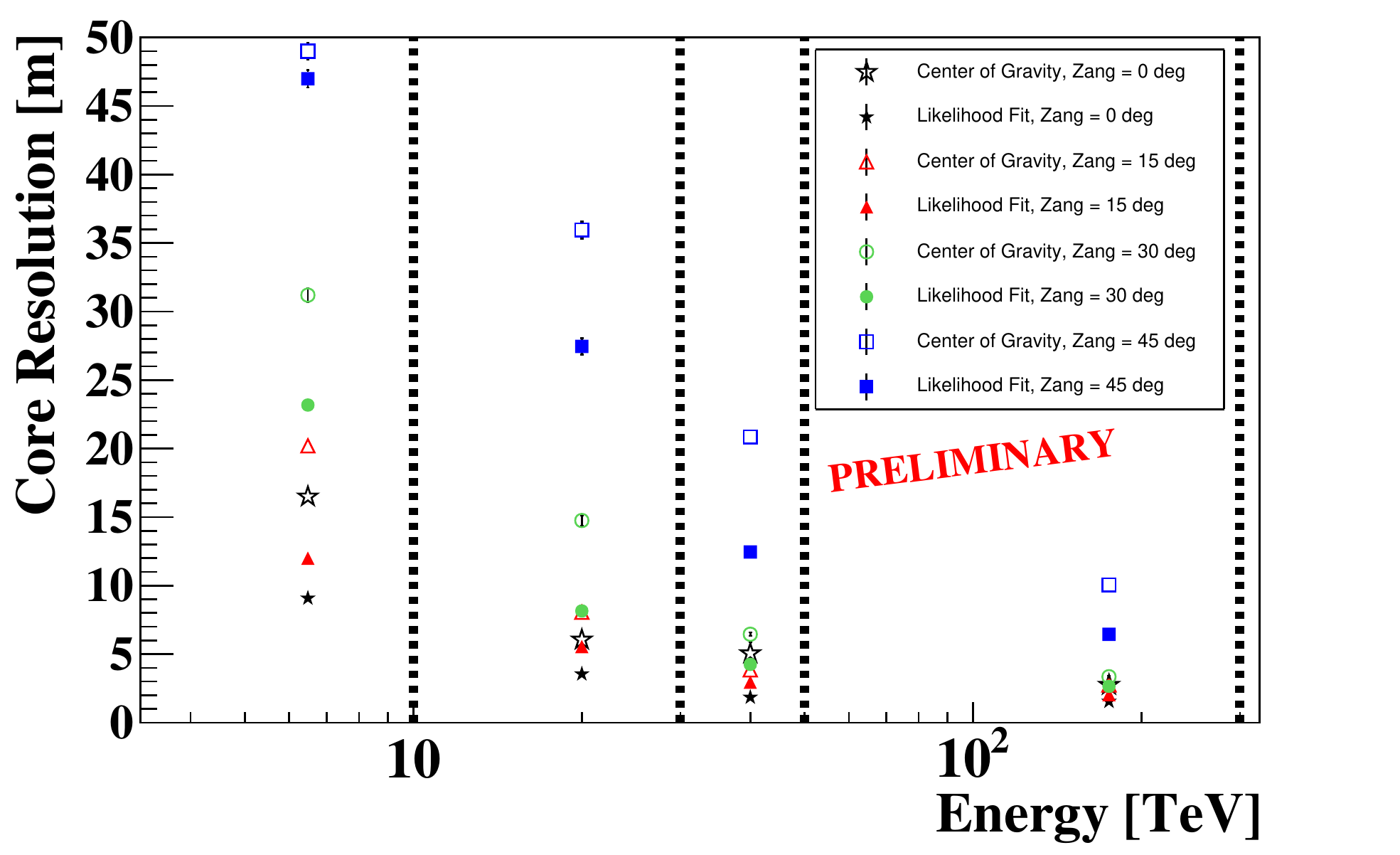}
\caption{The core resolution obtained with a likelihood fit in comparison with the center of gravity of the signal for different zenith angles (Zang). The vertical dashed lines represent the binning in the energy range. The points in each of these energy bins correspond to the 68\% containment of the core resolution distribution.}
\label{fig-4}       
\end{figure}

To constrain the core location of the multi-TeV $\gamma$-ray showers falling outside the main HAWC array a likelihood core fitter is being developed. In Figure~\ref{fig-4} we can see that a core resolution of < 10 m is achieved by just using the outriggers for energies > 10 TeV and for zenith angle up to 30$^{\circ}$. In addition, this likelihood method also constrains the shower energy and depth of the shower maximum. In the next step, this likelihood fit method for the outriggers will be merged with the one for the main array to ultimately improve the core resolution for multi-TeV showers.

\section{Current Status of the Outrigger Array}
\label{current status} 
The deployment of the outrigger array has already started. FALCON electronics is being used to take the data from the first set of outriggers installed at the HAWC site. Integration of the FALCON readout with the central DAQ is ongoing and will be finished soon. A complete outrigger array will be fully operational by the end of the next year.

\section*{Acknowledgements}
\label{ack}
We acknowledge the support from: the US National Science Foundation (NSF); the US Department of Energy Office of High-Energy Physics; the Laboratory Directed Research and Development (LDRD) program of Los Alamos National Laboratory; Consejo Nacional de Ciencia y Tecnologa (CONACyT), Mexico (grants 260378, 55155, 105666, 122331, 132197, 167281); Red de Fsica de Altas Energas, Mexico; DGAPA-UNAM (grants IG100414-3, IN108713, IN121309, IN115409, IN113612); VIEP-BUAP (grant 161-EXC-2011); the University of Wisconsin Alumni Research Foundation; the Institute of Geophysics, Planetary Physics, and Signatures at Los Alamos National Laboratory; the Luc Binette Foundation UNAM Postdoctoral Fellowship program.
%

\end{document}